\documentstyle[prd,tighten,aps]{revtex}
\begin{document}
\draft
\twocolumn[\hsize\textwidth\columnwidth\hsize\csname
@twocolumnfalse\endcsname
\title{Anisotropic cosmologies containing isotropic background
radiation}
\author{Saulo Carneiro$^{1,2}$ and Guillermo A. Mena Marug\'{a}n$^1$}
\address{$^1$ Instituto de Matem\'{a}ticas y F\'{\i}sica Fundamental,
C.S.I.C., Serrano 121, 28006 Madrid, Spain\\$^2$ Instituto de
F\'{\i}sica, Universidade Federal da Bahia, 40210-340, Salvador,
Bahia, Brazil}
\maketitle

\begin{abstract}
We present an anisotropic cosmological model based on a new exact
solution of Einstein equations. The matter content consists of an
anisotropic scalar field minimally coupled to gravity and of two
isotropic perfect fluids that represent dust matter and radiation.
The spacetime is described by a spatially homogeneous, Bianchi
type III metric with a conformal expansion. The model respects the
evolution of the scale factor predicted by standard cosmology, as
well as the isotropy of the cosmic microwave background.
Remarkably, the introduction of the scalar field, apart from
explaining the spacetime anisotropy, gives rise to an energy
density that is close to the critical density. As a consequence,
the model is quasiflat during the entire history of the universe.
Using these results, we are also able to construct approximate
solutions for shear-free cosmological models with rotation. We
finally carry out a quantitative discussion of the validity of
such solutions, showing that our approximations are acceptably
good if the angular velocity of the universe is within the
observational bounds derived from rotation of galaxies.
\end{abstract}

\vskip2pc]
\renewcommand{\thesection}{\Roman{section}}

\section{Introduction}

One of the best established facts in observational cosmology is
the isotropy of the cosmic microwave background (CMB) \cite{cobe}.
This high degree of isotropy explains the success of cosmological
perturbation theory \cite{cpt} in reproducing the spectrum of
anisotropies detected in the CMB \cite{Boom-Max}. The measurement
of these anisotropies, originated from primordial fluctuations,
has played a fundamental role in the advent of precision
cosmology, allowing for the first time the determination of
several cosmologi\-cal parameters and the rejection of a large
number of cosmological models \cite{BM2}.

The isotropy of the CMB, together with the apparent homogeneity
and isotropy of clustering matter, smeared out over scales of the
order of 100 Mpc \cite{hom}, provide the main experimental support
for the cosmological principle \cite{Weinberg}. The spatial
homogeneity and isotropy of the universe is incorporated in the
standard cosmological model by using the
Friedmann-Robertson-Walker (FRW) family of metrics to describe the
spacetime. It should be clear, nevertheless, that the cosmological
principle is actually a reasonable and fruitful hypothesis, rather
than a proven fact. In order to clarify this issue, at least from
a conceptual point of view, we want to show that it is possible to
construct spatially homogeneous cosmologies which are anisotropic
but still compatible with the observed isotropy of the background
radiation and the matter with strong clustering properties.

Our starting point will be a spatially homogeneous spacetime
metric of the Bianchi type III \cite{kram} subject to conformal
expansion. As shown in Ref. \cite{KO}, the condition that the
expansion be conformal is crucial for the isotropy of the CMB.
Part of the matter content will be given by a two-component
perfect fluid describing the (idealized) radiation and dust
matter that are present in our universe. By its own, this
perfect fluid cannot account for the anisotropy of the spacetime
\cite{EGS}. However, we will show that this problem can be
solved just by including an additional matter source consisting
in an anisotropic scalar field. In this way, we will be able to
construct an exact solution of Einstein equations with the
required properties.

In addition to their important role in inflationary models
\cite{inf} and Brans-Dicke cosmologies \cite{Weinberg,BD}, the use
of scalar fields in cosmology has received renewed attention
during recent years. The observed relation between luminosity
distance and redshift for type Ia supernovae (SNe Ia) has supplied
strong evidence in favor of an accelerated expansion of the
universe \cite{sne}. In order to explain this acceleration and fit
the SNe Ia data, cosmological models with a new matter component
have been proposed. This component, called quintessence, can be
modeled by a light scalar field with a self-interaction potential
and with minimal coupling to gravity \cite{quin}. Comparison of
the theoretical predictions with SNe Ia and CMB observations leads
to an estimate for the dark energy density of this quintessence
field that (roughly speaking) is of the order of magnitude of the
critical density \cite{sne,de}.

From this perspective, our approach to construct anisotropic
solutions can be considered as a new application of scalar fields
in cosmology. The anisotropic field that we will introduce does
not really model a quintessence component, because its presence
does not accelerate the expansion of the universe. Actually, as we
will see, our scalar field produces neither acceleration nor
deceleration, so that it can rather be regarded (apart from the
anisotropy) as the limit of a quintessence contribution when the
acceleration vanishes. Much more importantly, it turns out that
the dark energy density associated with this field is similar to
the critical density of the model at late times. In this sense,
our model suggests that the inclusion of an anisotropic scalar
field may provide a mechanism to generate quasiflat \cite{bama}
universes.

In the case of our anisotropic cosmological solution, the
quasiflatness can be rooted in the fact that the energy density of
the anisotro\-pic field is proportional to the inverse square of
the scale factor, which is precisely the type of dependence that
one would expect for the critical density at the final stages of
the expansion. The reason is that, in a Bianchi type III universe,
there exists a negative contribution of curvature to the energy
density, just like in an open FRW model, so that at large
cosmological times one would expect the scale factor to be
inversely proportional to the Hubble constant and, therefore, to
the square root of the critical density.

As we have said, the evolution of the aniso\-tropic solutions
that we will present is not accelerated, so that our
cosmological model cannot be considered realistic in this
respect. One could remedy this situation by including an
additional matter source, given by a quintessence field or a
cosmological constant. Instead of proceeding in that way, we
have preferred to keep the model as simple as possible, both to
isolate the cosmological consequences of the anisotropic scalar
field and to obtain a solution of Einstein equations in which
one can complete all calculations of cosmological parameters
using exact expressions.

Apart from clarifying the role that anisotropic scalar fields may
play in cosmology, we will also discuss in detail another possible
application of our exact anisotropic solution. This application
follows from the fact that the spacetime metric of our model is
just the vanishing-rotation member of a family of spatially
homogeneous, rotating and shear-free metrics that were studied by
Korotki\v{\i} and Obukhov (KO) \cite{KO,Obukhov}. This family is
an expanding version of a class of G\"{o}del-like stationary metrics
analyzed by Rebou\c{c}as and Tiomno (RT) \cite{RT}. From now on, we
will refer to them as the RTKO metrics. In the limit of small
rotation, it is not difficult to employ our exact anisotropic
solution to construct approximate solutions for the rotating RTKO
cosmologies and supply them with a physically acceptable matter
content.

We will also carry out a quantitative analysis of the validity of
our approximations when a small rotation is present. In
particular, we will find a bound on the angular velocity of the
universe guaranteeing that our approximate solutions cannot be
distinguished from the unknown exact ones, at least as far as the
energy-momentum tensor is concerned.

The RTKO metrics share in fact most of the good properties
presented by our exact anisotropic solution with vanishing
rotation. For instance, they are all of the Bianchi type III and
possess a conformal Killing vector field (CKVF). The existence
of a vector of this kind parallel to the four-velocity of dust
matter and radiation is known to be the necessary and sufficient
condition for the absence of parallax effects \cite{para} and
turns out to guarantee the isotropy of the CMB \cite{KO}. In
addition, the spacetime does not contain closed timelike curves
(CTC's) unless the rotation is considerably large \cite{KO}. On
the other hand, it has been shown that the RTKO metrics
reproduce the open FRW metric in the limit of small rotation and
nearby distances \cite{PRD}. All these properties clearly make
of the RTKO metrics natural candidates to describe anisotropic
cosmological scenarios with rotation.

The rest of the paper is organized as follows. Since the spacetime
of the exact cosmological solution that we will construct belongs
to the RTKO family, we will first analyze the most relevant
properties of these geometries and obtain their Einstein equations
in Sec. II. In Sec. III we present our exact anisotropic solution
with vanishing rotation, and discuss the corresponding cosmology
in Sec. IV. Based on the case of zero angular velocity, we find in
Sec. V approximate RTKO solutions with rotation. We also study
their cosmological parameters and show that the quasiflatness of
the cosmological model persists in the presence of rotation. By
comparing the Einstein tensor of our approximate solutions with
the energy-momentum tensor of its assumed matter content, we
derive in Sec. VI a bound on the angular velocity of the universe
ensuring that the relative errors committed in the Einstein
equations are small. The conclusions of our work are included in
Sec. VII. Finally, an appendix containing some calculations is
added.

\section{The RTKO  metrics}
\setcounter{equation}{0}
\renewcommand{\theequation}{\arabic{section}.\arabic{equation}}

The RTKO metrics are described by the line element
\begin{equation}
\label{KO}
ds^2\!=\!a^2(\eta)\!\left[-(d\eta+le^xdy)^2\!+\!
dx^2+e^{2x}dy^2+dz^2\right],
\end{equation}
where $\eta$ is the conformal time and $x,y,z$ are the spatial
coordinates, all of them assumed to run over the real axis. The
parameter $l$, on the other hand, is a constant that can be
restricted to be non-negative without loss of generality. From
now on, we call it the rotation parameter. The above spacetime
does not contain CTC's if and only if $l$ belongs to the
interval $[0,1)$, since it is only then that the metric induced
on the sections of constant time is positive definite
\cite{KO,RT}. In the following, we will restrict our
considerations to this causal sector, $0\leq l<1$.

We will employ the following notation. Greek letters will denote
spacetime indices, and the indices $\{0,1,2,3\}$ will designate,
respectively, the coordinates $\{\eta,x,y,z\}$. In addition, we
adopt units such that $8\pi G=c=1$, $G$ being Newton constant.

Metric (\ref{KO}) is a spatially homogeneous, Bianchi type III
metric, with three Killing vector fields given by
\begin{equation}
\xi_{(1)}=\partial_x-y\partial_y,\hspace*{.8cm}
\xi_{(2)}=\partial_y,\hspace*{.8cm}
\xi_{(3)}=\partial_z.\end{equation} The metric possesses also a
CKVF, namely, $\xi^{\mu}_{C}=\delta^{\mu}_{0}$. Korotki\v{\i} and
Obukhov have proved that, assuming comoving radiation and a
comoving observer, the existence of a CKVF guarantees the isotropy
of the detected CMB, with the radiation temperature falling with
the inverse of the scale factor, and ensures that the redshift of
the light coming from astrophysical objects does not depend
(explicitly) on the spatial positions of the source and the
receiver, but only on the emission and observation times
\cite{KO,Obukhov}. Besides, the CKVF prevents the appearance of
parallax effects \cite{para}. Furthermore, in these spacetimes the
shear tensor vanishes for comoving observers (with four-velocity
equal to $u^{\mu}=\delta^{\mu}_0/a$), whereas their rotation
tensor $\omega_{\mu\nu}$ is different from zero \cite{kram,KO}. In
particular, their angular velocity is
$\omega=\sqrt{\omega_{\mu\nu}\omega^{\mu\nu}/2}=l/(2a)$.

With the help of the coordinate transformation
\begin{eqnarray}
\label{coorx} e^x&=&\cosh{r}+\cos{\phi}\sinh{r},\\
\label{coory}ye^x&=&\sin{\phi}\sinh{r},\\
\label{cooret}\eta&=&\tilde{\eta}-l\phi+
2l\arctan{\!\left(e^{-r}\tan{\frac{\phi}{2}}\right)},
\end{eqnarray}
one can write metric (\ref{KO}) in the cylindrical form
\cite{RT}
\begin{eqnarray}
\label{metrica} ds^2&=&-a^2(\eta)
\left[d\tilde{\eta}+2l\sinh^2{\!\left(\frac{r}{2}\right)}d\phi
\right]^2\nonumber\\
&&+a^2(\eta)\left[dr^2+\sinh^2{\!r}\; d\phi^2+dz^2\right].
\end{eqnarray}
Here, $\tilde{\eta}$ is the new real time, $r$ is non-negative,
and $\phi$ is an angular coordinate. Note that the scale factor is
not constant on the sections of constant time $\tilde{\eta}$
unless the rotation parameter vanishes [or $a(\eta)$ is a constant
number], because $a$ depends on the radial and angular coordinates
$r$ and $\phi$ when $l\neq 0$. On the other hand, using the change
of coordinates (\ref{coorx}) and (\ref{coory}), one can easily
check that the sections of constant time $\eta$ are the direct
product of a real line and a two-dimensional pseudosphere.

Let us finally consider the associated Einstein equations. For
the diagonal components of the Einstein and energy-momentum
tensors, one obtains
\begin{eqnarray}
\label{E0}T^0_0 a^4 &=&
\left(1-\frac{3l^2}{4}\right)a^2-3(1-l^2)\dot{a}^2,\\
\label{E1}
T^1_1a^4 = T^2_2a^4 &=&
\frac{l^2}{4}a^2+(1-l^2)\dot{a}^2-2(1-l^2)a\ddot{a},\\
\label{E3}
T^3_3a^2&=&T^1_1a^2+1-\frac{l^2}{2},
\end{eqnarray}
where the overdot denotes the derivative with respect to the
conformal time $\eta$. For the non-diagonal components, on the
other hand, the result is
\begin{eqnarray}
\label{nd} T^1_0 a^4&=&l^2 a\dot{a},\nonumber \\
T^2_0 a^4&=&2le^{-x} (a\ddot{a}-2\dot{a}^2),\nonumber \\ T^2_1
a^4 &=&-le^{-x}a\dot{a},\nonumber \\ T^0_2 a^4&=&(1-l^2)le^x
a^2,\nonumber\\ T^1_2 a^4&=&-(1-l^2)le^xa\dot{a}.
\end{eqnarray}
The remaining components of the energy-momentum tensor must be
identically zero. As far as we know, no physically admissible
matter source has been proposed up to date leading to a solution
of the above Einstein equations when the scale factor is not
constant. Hence, no explicit RTKO cosmological model has been
constructed so far. In the next section, we will present an
exact solution for the case of vanishing rotation that has an
acceptable energy-momentum tensor. This solution represents an
anisotropic universe in continuous expansion.

\section{The anisotropic solution}
\setcounter{equation}{0}
\renewcommand{\theequation}{\arabic{section}.\arabic{equation}}

We will now restrict our attention to the RTKO metric obtained
when the rotation parameter $l$ vanishes. In the absence of
rotation, the Einstein equations require the energy-momentum
tensor to be diagonal. The diagonal components must satisfy
\begin{eqnarray}
\label{E0d}
\epsilon a^4 &=& 3\dot{a}^2-a^2,\\
\label{E1d}
p_1a^4 = p_2a^4& =& \dot{a}^2-2a\ddot{a},\\
\label{E3d}
p_3a^2&=&p_1a^2+1.
\end{eqnarray}
Here, we have adopted the notation $\epsilon\equiv -T^{0}_{0}$ for
the energy density and $p_i\equiv T^i_i$ ($i=1,2$, or 3) for the
principal pressures of the system.

Let us start by assuming that $\epsilon a^2$ vanishes in the limit
of infinite scale factor, as would happen if the matter content
consisted exclusively of dust and radiation
\cite{Weinberg,Landau}. From Eq. (\ref{E0d}) one then easily sees
that $\dot{a}^2/a^2$ must tend to $1/3$ when $a\rightarrow\infty$.
Note also that this equation ensures that $a$ increases
unboundedly with the conformal time, provided that $\dot{a}$ is
initially positive. In addition, supposing that $\epsilon a^2$ is
a smooth function of the scale factor, the condition $\epsilon
a^2\rightarrow 0$ implies that $(d\epsilon/da)a^3$ vanishes when
$a$ becomes infinitely large. Using these facts and taking the
time derivative of Eq. (\ref{E0d}), it follows that the quotient
$\ddot{a}/a$ must also tend to $1/3$ when $a$ approaches infinity.
Employing now Eq. (\ref{E1d}), one concludes that $p_1a^2$ has a
finite limit when $a\rightarrow\infty$. As a result, the dominant
energy condition \cite{HE} is violated during the evolution,
because for sufficiently large scale factors the pressure $p_1$
becomes larger than the energy density.

Therefore, if we want to reach an acceptable solution of the
Einstein equations that respects the energy conditions, we must
include matter sources whose energy density does not fall faster
than $1/a^2$ when the scale factor expands to infinity.
Probably, the simplest way to do this is by introducing an
anisotropic massless scalar field minimally coupled to gravity.
As we will see below, the corresponding energy density satisfies
precisely the minimal requirement of being proportional to the
inverse square of the scale factor. Furthermore, the inclusion
of such a scalar field will actually suffice to explain all the
anisotropies of the model, allowing the rest of the matter
content to be isotropic.

In curved spacetime, a massless minimally coupled scalar field
satisfies the equation
\begin{equation}
\label{SF}
\Phi_{;\mu\nu}g^{\mu\nu}=\frac{1}{\sqrt{-g}} \left(
\sqrt{-g} \Phi_{,\mu} g^{\mu \nu} \right)_{,\nu} =
0,
\end{equation}
whereas its energy-momentum tensor has the form
\begin{equation}
\label{SFT}
T^{\nu}_{\mu}=\Phi_{,\mu}\Phi_{,\sigma}g^{\sigma\nu}
-\frac{1}{2}\Phi_{,\sigma}\Phi_{,\rho} g^{\sigma
\rho}\delta^{\nu}_{\mu}.\end{equation}
Here, $g$ and $g^{\mu\nu}$ are the determinant and the inverse
of the four-metric, the semicolon denotes covariant derivative,
and $\delta^{\nu}_{\mu}$ is the Kronecker delta.

Let us then consider an anisotropic scalar field given by
$\Phi=Cz$, with $C$ being a constant. This kind of source for
the RTKO metrics was already suggested by Rebou\c{c}as and Tiomno in
a stationary context with rotation \cite{RT}. It is easily
checked that Eq. (\ref{SF}) is in fact satisfied by our field in
any of the spacetimes (\ref{KO}). Besides, from Eq. (\ref{SFT}),
the solution $\Phi=Cz$ has a diagonal energy-momentum tensor,
with the following energy density and principal pressures:
\begin{equation}\label{EPSF}
\epsilon^{(s)}=p_3^{(s)}=-p_2^{(s)}=-p_1^{(s)}=\frac{C^2}{2a^2}.
\end{equation}
Here, the superindex ${(s)}$ refers to the contribution of the
scalar field.

Note that the corresponding energy density falls with $a^2$, as we
had anticipated. In this respect, it is interesting to note that
such a kind of decay for the energy density is also expected on
the basis of quantum cosmo-logy arguments and might even provide a
way to solve the cosmological constant problem \cite{WJJ}.

In addition, the principal pressures are now aniso\-tropic. Using
this property, it is actually very simple to remove any trace of
anisotropy from the Einstein equations of our spacetime. Defining
$\epsilon\equiv\bar{\epsilon}+\epsilon^{(s)}$ and
$p_i\equiv\bar{p}_i+p_i^{(s)}$, we see from Eq. (\ref{E3d}) that
the anisotropic contributions of the model can be absorbed in the
scalar field just by imposing that $C^2=1$. Since the orientation
of $z$ can be inverted at will (producing an apparent flip of sign
in the constant $C$), we will fix from now on $\Phi=z$.

Equations (\ref{E0d})-(\ref{E3d}) become then
\begin{eqnarray}
\label{bare}
\bar{\epsilon} a^4 &=&
3\dot{a}^2-\frac{3}{2}a^2,\\
\label{barp}
\bar{p}a^4 &=& \dot{a}^2-2a\ddot{a}+\frac{a^2}{2},
\end{eqnarray}
where $\bar{p}=\bar{p}_i$ for any $i=1,2$ or 3. Remarkably,
these are exactly the Einstein equations of an open FRW model
with curvature parameter $\kappa$ equal to $-1/2$
\cite{Weinberg}. Equivalently, they can be written as the
equations of the standard FRW model with $\kappa=-1$ under the
scaling:
\begin{eqnarray}
\eta&\equiv&\sqrt{2}\eta_{F}\nonumber\\
\label{FRW}
\sqrt{2}\;a\!\left(\eta=\sqrt{2}\eta_{F}\right)&\equiv &a_{F}(\eta_{F}).
\end{eqnarray} Here, the subindex $F$ denotes the conformal
time and scale factor of the open FRW cosmology. Notice that
these relations imply that the cosmological time of our model
coincides with that of the standard FRW spacetime, because
$ad\eta=a_{F}d\eta_{F}$.

From the above comments, it should be clear that the evolution of
the scale factor in our model reproduces the expansion found in an
open FRW cosmology, except for some qualitatively irrelevant
scalings by factors of the order of the unity. Owing to this fact,
and leaving aside the anisotropy of the sections of constant time,
the cosmological solution that we will construct leads essentially
to the same history of the universe as a standard open FRW
scenario, at least during the epoch in which the scalar field has
a negligible contribution to the energy and pressure of the
system. Furthermore, regarding the anisotropy of the spatial
sections, we recall that our spacetime metric can be written in
the form (\ref{metrica}) with $l=0$. In fact, since such metric
reduces to an open FRW metric in the limit of nearby distances
$r\ll 1$ \cite{PRD}, no differences should be expected in physical
processes or observations which do not involve distant regions.

We are now in an adequate position to obtain the solution of the
Einstein equations that we were seeking. In addition to the
anisotropic scalar field, we suppose that the matter content is
given by radiation and dust, as it is usually done in standard FRW
cosmology. We will describe these matter sources by a
two-component perfect fluid, with comoving four-velocity $u^{\mu}=
\delta^{\mu}_{0}/a$. The assumption that the radiation present in
the system adopts the form of a comoving perfect fluid, together
with the properties of the RTKO metrics \cite{KO}, guarantees that
the CMB of the model is isotropic. Similarly, the fact that the
dust matter can be treated as a comoving perfect fluid ensures the
applicability of Hubble law (in the leading-order approximation)
to any kind of radiation that could be emitted by dust particles,
since the radiation frequency varies then just like the inverse of
the scale factor\cite{KO}. For such a matter content, the
expression of the energy and pressure that appear in Eqs.
(\ref{bare}) and (\ref{barp}) are
\begin{equation}\label{barep}
\bar{\epsilon}=\frac{A^2}{a^4}+\frac{D}{a^3},\hspace*{1.2cm}
\bar{p}=\frac{A^2}{3a^4},\end{equation}
where $A$ and $D$ are two non-negative constants. The first term
on the right-hand side of these equations corresponds to the
radiation component, whereas the dust matter contributes only to
the energy density \cite{Weinberg,Landau}.

With the above energy and pressure, Eq. (\ref{bare}) turns out to
be a first integral of Eq. (\ref{barp}), and admits a unique
increasing solution that vanishes at $\eta=0$. The exact solution
is given explicitly by
\begin{equation}
\label{sol}
a=\frac{D}{3}\left[\cosh\left(\frac{\eta}{\sqrt{2}}\right)-1\right]+
\sqrt{\frac{2}{3}}A\sinh\left(\frac{\eta}{\sqrt{2}}\right).
\end{equation}
This expression can be inverted in $\eta\geq 0$, obtaining
\begin{equation} \label{confo}
\eta = \sqrt{2}
\ln\left[\frac{3a+D+\sqrt{9a^2+6Da+6A^2}}{D+\sqrt{6}A}\;
\right].
\end{equation}
On the other hand, integrating $dt=ad\eta$, we arrive at the
following expression for the cosmological time:
\begin{eqnarray}\label{t}
t&=&\frac{D}{3}\left[-\eta+\sqrt{2}\sinh\left(\frac{\eta}{\sqrt{2}}
\right)\right]\nonumber \\
&&+\frac{2A}{\sqrt{3}}
\left[\cosh\left(\frac{\eta}{\sqrt{2}}\right)-1\right].
\end{eqnarray}
From the last two formulas, one can also calculate $t$ as a
function of the scale factor.

\section{The anisotropic cosmological model}
\setcounter{equation}{0}
\renewcommand{\theequation}{\arabic{section}.\arabic{equation}}

In the preceding section, we have constructed an exact solution of
Einstein equations that describes an expanding universe containing
an anisotropic massless scalar field and a comoving perfect fluid
composed of radiation and dust matter. The spacetime metric is
given by the element of the family (\ref{KO}) with vanishing
rotation. As a consequence of the properties of the RTKO metrics,
the CMB of the model is isotropic and the redshift of the
radiation emitted by the comoving dust depends only on the
emission and observation times \cite{KO}. We have also seen that
our anisotropic metric coincides with the metric of an open FRW
universe in the limit of nearby distances. Moreover, the conformal
expansion of our solution reproduces (apart from some trivial
scalings) the evolution encountered in a standard open FRW
cosmo-logy with matter content formed exclusively by isotropic
dust and radiation. Hence, the history of the universe in our
anisotropic model parallels that of an open FRW solution, at least
as far as the scalar field does not supply the dominant
contribution to the energy-momentum tensor.

Like in the analogue FRW cosmology with energy density and
pressure given by $\bar{\epsilon}$ and $\bar{p}$, the radiation
dominated era of our anisotropic model corresponds to the epoch
with $0\leq a\leq A^2/D$. At small times, the universe expands
from an initial singularity following exactly the same evolution
law as in standard FRW cosmology \cite{Weinberg,Landau}, namely,
\begin{equation}\label{small}
a=\frac{A\eta}{\sqrt{3}}=\sqrt{\frac{2At}{\sqrt{3}}}.
\end{equation}
This behavior can be easily obtained from Eq. (\ref{sol}) in the
region $\eta\ll 1$. As a particular consequence, the Hubble
parameter and the energy density adopt, at the initial stages of
the expansion, the expressions $H\equiv\dot{a}/a^2=1/(2t)$ and
$\bar{\epsilon}=3/(4t^2)$, which coincide with the result of the
standard model in the radiation era. In particular, it follows
that the initial relative energy density is
$\Omega=\bar{\epsilon}/(3H^2)=1$.

When $a$ increases beyond $A^2/D$, the dust component starts to
supply the major contribution to the energy density and the
universe enters a dust dominated era with an evolution of the
scale factor similar to that presented in an open FRW cosmology.
Such era ends when the energy of the anisotropic scalar field
becomes the most important matter component. This occurs when
$\epsilon^{(s)}$ equals the dust energy density, i.e., when
$a=2D$. We assume that $A\ll \sqrt{2}D$, so that there exists a
sufficiently large epoch $A^2/D\leq a\leq 2D$ dominated by
matter with strong clustering properties. As far as $a<2D$, the
contribution of the anisotropic scalar field is subdominant, and
the model leads essentially to the same cosmological predictions
as an open FRW model.

For scale factors larger than $2D$, the anisotropic scalar field
dominates the evolution. The expansion is then of the
approximate form
\begin{equation} \label{linear}
a=\left(\frac{A}{\sqrt{6}}+\frac{D}{6}\right)\exp{\left(
\frac{\eta}{\sqrt{2}}\right)}=
\frac{t}{\sqrt{2}},
\end{equation}
as one can check from Eq. (\ref{sol}) by analyzing the sector of
large times. Note that this evolution is linear in the
cosmological time, like at the final stages of an open FRW model.
This was in fact expected, because the time dependence of the
scale factor must always be similar to that of an open FRW
universe without scalar field, as we showed in the preceding
section. In the limit $\eta\rightarrow\infty$, the Hubble
parameter displays then the behavior $H=1/t$, and the energy
density is $\epsilon^{(s)}=H^2$. Hence, at large times, the
relative energy density becomes $\Omega=1/3$.

Actually, from Eq. (\ref{sol}) we can derive the exact
expressions of the Hubble parameter, the deceleration parameter
$q$, and the relative energy density \cite{Weinberg} at all
times of the evolution. We get
\begin{eqnarray} \label{H}
H &\equiv &\frac{\dot{a}}{a^2} = \sqrt{ \frac{3a^2+2Da+
2A^2}{6a^4}},\\ \label{O}
\Omega &\equiv& \frac{\epsilon}{3H^2} =
\frac{a^2+2Da+2A^2}{3a^2+ 2Da+2A^2},\\
\label{q} q &\equiv& 1 -\frac{a\ddot{a}}{\dot{a}^2}=
\frac{Da+2A^2}{3a^2+2Da+2A^2}>0,
\end{eqnarray}
where we have used that $\epsilon=\epsilon^{(s)}+\bar{\epsilon}$
and employed Eqs. (\ref{EPSF}) and (\ref{barep}). We recall that
the parameter $q$ is positive when the expansion decelerates.

In the limits $a \rightarrow 0$ and $a\rightarrow\infty$ (i.e.,
when $\eta$ tends to zero and infinity, respectively), we recover
from these equations the behavior discussed above for $H$ and
$\Omega$. Furthermore, it is not difficult to prove that Eq.
(\ref{O}) defines a strictly decreasing function of the scale
factor, $\Omega(a)$. Since the universe is always expanding in our
solution, we conclude that the relative energy density of our
model suffers a continuous decrease from its initial unit value at
the big-bang singularity, reaching the asymptotic lower bound of
one-third in the limit of large times. In this way, the
contribution of the anisotropic scalar field guarantees that the
energy density of the model is of the order of the critical one
during the whole evolution, leading to a quasiflat universe.

Obviously, the model is not fully realistic; in particular, the
positivity of Eq. (\ref{q}) means that the expansion decelerates
in our solution, contradicting the present observations of SNe Ia
\cite{sne}. The result $q>0$ can be easily understood on the basis
of our matter content: as we have seen, the scalar field leads to
a uniform expansion, linear in the cosmological time, whereas the
presence of radiation and dust decelerates the expansion. Note,
however, that the deceleration is similar to that found in a
standard open FRW cosmology without cosmological constant and
quintessence fields. This follows from the fact that the
deceleration parameter $q$ reflects only the time dependence of
the scale factor, and this dependence coincides in our solution
and in an open FRW model.

In order to attain an accelerated expansion in our anisotropic
scenario, we could simply add a positive cosmological constant
$\Lambda$ to the matter content. Indeed, it is easy to check that,
for the epoch in which $\Lambda$ dominates the energy density, Eq.
(\ref{E0d}) would lead to an exponential expansion. Like in
standard FRW cosmology, however, we have preferred to analyze here
the case without cosmological constant (or quintessence) because
in this way we can obtain an explicit solution that allows us to
perform all calculations to conclusion. In addition, the inclusion
of other matter sources would have prevented us from clearly
isolating the consequences of the anisotropic scalar field.

In order to estimate the values of the parameters $A$ and $D$
and the present values of $a$, $t$, $q$, and $\Omega$ in our
model, we can proceed as follows. From Eq. (\ref{H}) we get
\begin{equation}
a_0=\sqrt{\frac{3}{6H_0^2-2\bar{\epsilon}_r-2\bar{\epsilon}_d}},
\end{equation}
where the subindex 0 means evaluation at the present time, and the
sub\-indices $r$ and $d$ denote the radiation and dust components
of the matter content. In addition, if $\Omega_d$ is the
contribution of dust matter to the relative energy density and
$Z_{eq}$ is the redshift corresponding to the equilibrium between
dust and radiation, we have that $\bar{\epsilon}_d=3\Omega_dH_0^2$
and $\bar{\epsilon}_r=\bar{\epsilon}_d(1+Z_{eq})^{-1}$. Finally,
$A^2=\bar{\epsilon}_ra_0^4$ and $D=\bar{\epsilon}_da_0^3$. With
these values and formulas (\ref{O}) and (\ref{q}), we can also
determine the quantities $q_0$ and $\Omega_0$. Using
(approximately) the values of the concordance model \cite{conc}
for the present Hubble parameter and relative energy density of
pressureless matter, $H_0=65$ km/(sMp) and $\Omega_d=0.35$, as
well as $Z_{eq}+1=5000$, we obtain that $A=1.6\times10^{24}$ m,
$D=1.0\times10^{26}$ m, $a_0=1.2\times10^{26}$ m, $t_0=12$ Gyr,
$q_0=0.18$, and $\Omega_0=0.57$.

From these estimates, we see that the assumption $A\ll \sqrt{2}D$
is actually satisfied in our solution. The dust era corresponds to
the interval of scale factors $2.5\times10^{22}\;{\rm m}\leq a\leq
2.0\times 10^{26}$ m, which is large enough for structure
formation and contains the present period of the evolution. We
also see that the equilibrium between dust matter and the
anisotropic scalar field would be reached when
$a=2D=2.0\times10^{26}$ m, a value of the scale factor that is
only slightly larger than the present one. Thus, in our
cosmological model, we would be almost at the end of the dust
dominated epoch.

It is worth noting that, although the CMB of the model is
isotropic and the redshift of the radiation emitted by dust
particles depends only on the value of the scale factor at the
moment of emission, and not on the spatial position of the source,
the fact that the metric is anisotropic implies that the distance
to astrophysical objects with identical redshift varies with the
direction of observation. One might then worry about the
compatibility of this anisotropy with the available data about
extra-galactic sources at high redshift, e.g. with the apparent
isotropy detected in the Hubble diagram for SNe Ia at redshifts of
order unity. In order to discuss this issue, let us consider the
angular diameter distance \cite{Weinberg}, which can be defined by
the relation $dA_e=r_a^2\;d\Omega_0$ \cite{Obukhov2}. Here, $dA_e$
is the (infinitesimal) intrinsic perpendicular area of the source,
which subtends the solid angle $d\Omega_0$ at the origin where,
using the homogeneity of the spacetime, we locate the receiver
\cite{Obukhov2,KS}. The luminosity distance is then
$r_l=r_a(1+Z)^2$, with $Z$ being the redshift of the source
\cite{KS}. Hence, one only has to care about the anisotropies that
appear in $r_a$. Using the expressions given in Ref.
\cite{Obukhov2} (or just applying the formulas of Ref.
\cite{MCET}), it is possible to show that
\begin{equation}\label{RA}
r_a^2=a^2(\eta_e)\;(\eta_0-\eta_e)^2\;Y[\sin{\theta}(\eta_0-\eta_e)],
\end{equation}
where $Y(u)\equiv \sinh{u}/u$, $\theta\in [0,\pi]$ is the angle
formed by the line of sight and the $z$ axis, and $\eta_e$ is the
conformal time of emission. As anticipated, $r_a$ depends on the
direction of observation and, for fixed $Z$ (and present time
$\eta_0$), its maximum $r_M$ and minimum $r_m$ are reached when
$\sin{\theta}$ equals the unity or tends to zero, respectively.
The magnitude of the relative variation of $r_a$ on the celestial
sphere can be described with the quantity
$\varepsilon_a=(r_M-r_m)/r_m$. Employing Eqs. (\ref{confo}),
(\ref{RA}), and $1+Z=a_0/a(\eta_e)$, it is straightforward to see
that $\varepsilon_a$ increases with $Z$. More importantly,
substituting the values of the constants $A$, $D$, and $a_0$
obtained above, one can check that the relative variation of the
angular dia-meter distance is only of the order of 5$\%$ for
$Z=1$, while for $Z=2$ $\varepsilon_a$ is close to 10$\%$. These
variations do not seem to conflict with the observational data,
and do not dominate over the systematic and statistical
uncertainties, evolution effects, and experimental errors that are
present in the determination of astronomical distances.

Finally, let us point out that the age of the universe in our
model ($t_0=12$ Gyr), although very close, is still beyond the
lower bounds obtained from radioactive dating of stars \cite{age}
or studies of globular clusters \cite{age2}. These results show
that (except for the absence of acceleration and the corresponding
quintessence contribution to the relative energy density) our
anisotropic model is at least compatible with the main features of
modern standard cosmology.

\section{Approximate rotating solutions}
\setcounter{equation}{0}
\renewcommand{\theequation}{\arabic{section}.\arabic{equation}}

In this section, we will present a generalization of the solution
(\ref{sol}) for a non-vanishing rotation parameter, $l\neq0$. We
will assume the same matter content as in the absence of rotation,
namely, a two-component perfect fluid, formed by radiation and
dust, and an anisotropic scalar field $\Phi=z$ minimally coupled
to gravity. For small values of the parameter $l$, we will see
that the RTKO metric that we will obtain can be regarded as an
approximate solution of the Einstein equations. In this way, one
can construct an approximate cosmological model describing the
expansion of a rotating anisotropic universe which contains
isotropic background radiation. Actually, supposing that $l$ is
sufficiently small, the inclusion of rotation produces only small
corrections in the cosmological model constructed in Sec. IV. As a
consequence, our approximate solutions will lead to a similar
cosmology, both qualitatively (apart from the existence of an
angular velocity) and quantitatively.

The energy-momentum tensor will have the form
\begin{equation}
\label{EMT}
T^{\nu}_{\mu}=(\bar{p}+\bar{\epsilon})u^{\nu}u_{\mu}+\bar{p}
\delta^{\nu}_{\mu}+(T^{(s)})^{\nu}_{\mu},
\end{equation}
where $u^{\mu}$ is the four-velocity of the two-component fluid,
its energy density and pressure are given in Eq. (\ref{barep}),
and $T^{(s)}$ denotes the energy-momentum tensor of the
anisotropic scalar field. The components of this diagonal tensor
appear in Eq. (\ref{EPSF}) (with $C=1$). The parameters $A$ and
$D$, which determine the energy density, are assumed to be
exactly the same as in the solution with vanishing rotation.
Like in that case, we also consider comoving perfect fluids with
$u^{\mu}=\delta^{\mu}_0/a$.

Using the general RTKO non-diagonal metric (\ref{KO}), we obtain
the covariant four-velocity
$u_{\mu}=-a(\delta^0_{\mu}+le^x\delta^2_{\mu})$. Then, from our
definition (\ref{EMT}), we see that the diagonal components of
the energy-momentum tensor are (formally) the same as in our
solution with zero angular velocity, whereas all the
non-diagonal components vanish except $T^0_2$. This last
component takes the expression
\begin{equation}\label{T02}
T^0_2=-le^x\left(\frac{4A^2}{3a^4}+\frac{D}{a^3}\right).
\end{equation}

Let us first consider the diagonal time component (\ref{E0}) of
the Einstein equations. When the rotation parameter does not
vanish, this equation has the following solution for our value of
the energy density:
\begin{eqnarray}
\label{apsol}
a&=&\frac{D}{3X_l}\left[\cosh{\left(\sqrt{\frac{X_l}{2Y_l}}\;
\eta\right)}-1\right]\nonumber\\
&&+\sqrt{\frac{2}{3X_l}}A
\sinh{\left(\sqrt{\frac{X_l}{2Y_l}}\;\eta\right)},
\end{eqnarray}
where we have introduced the definitions
\begin{equation}\label{XYL}
X_l\equiv1-\frac{l^2}{2},\hspace*{1.2cm}Y_l\equiv 1-l^2.
\end{equation}
The above scale factor increases with the conformal time in
$\eta\geq 0$ and vanishes at $\eta=0$. In addition, it
reproduces Eq. (\ref{sol}) when $l$ vanishes. Note also that,
since we have imposed that $l\in[0,1)$, the ranges of $X_l$ and
$Y_l$ are, respectively, $(1/2,1]$ and $(0,1]$.

Substituting the above time dependence of the scale factor and the
expression of the energy-momentum tensor in Eqs. (\ref{E1}) and
(\ref{E3}), it is easy to check that the Einstein equation
$G^3_3=T^3_3$ is satisfied exactly; however, the other diagonal
spatial components of the Einstein tensor differ by a term
$l^2/(2a^2)$ from their assumed va-lues. In other words,
\begin{equation}
G^1_1-T^1_1=G^2_2-T^2_2=\frac{l^2}{2a^2}.
\end{equation}
Concerning the non-diagonal components (\ref{nd}) of the Einstein
equations, it is not difficult to prove using Eq. (\ref{T02})
that, when $l$ is small, the Einstein tensor of the analyzed RTKO
metric provides an approximate solution up to terms of the order
of $l^2$ for $G^1_0=T^1_0$ and of order $l$ for the rest of
equations. Therefore, we conclude that the difference between the
components of the energy-momentum tensor of our system and those
of the Einstein tensor of the metric (\ref{KO}) and (\ref{apsol})
vanish at least as fast as $l$ when $l\rightarrow 0$, and become,
in general, negligible when the rotation parameter is small. In
Sec. VI we will use this fact to set an upper bound to the global
angular velocity in order to ensure that the relative error
committed in the energy-momentum tensor with our approximation is
smaller than a certain quantity.

Let us now analyze the behavior of our approximate cosmological
solutions with rotation. Inverting relation (\ref{apsol}), we
obtain the conformal time
\begin{equation} \label{conformal}
\eta\! =\! \sqrt{\!\frac{2Y_l}{X_l}}
\ln\!{\left[\frac{3X_la+D+\sqrt{9X_l^2a^2+6X_lDa+6X_lA^2}}
{D+\sqrt{6X_l}\;A}\right]},
\end{equation}
and, integrating $dt=ad\eta$, we get the following expression
for the cosmological time:
\begin{eqnarray}
t&=&\frac{D}{3X_l}\left[-\eta+\sqrt{\frac{2Y_l}{X_l}}
\sinh{\left(\sqrt{\frac{X_l}{2Y_l}}\;\eta\right)}\right]\nonumber
\\&+&\frac{2A}{X_l}\sqrt{\frac{Y_l}{3}}
\left[\cosh{\left(\sqrt{\frac{X_l}{2Y_l}}\;\eta\right)}-1\right].
\end{eqnarray}

From relation (\ref{apsol}), one can also derive the Hubble
parameter, the deceleration parameter, and the relative energy
density of our approximate solutions:
\begin{eqnarray} \label{HR}
H &=&\sqrt{\frac{3X_la^2+2Da+2A^2}{6Y_la^4}},\\\label{OR}
\Omega &=&\frac{Y_l(a^2+2Da+2A^2)}{3X_la^2+2Da+2A^2},\\
\label{qR}
q &=&\frac{Da+2A^2}{3X_la^2+2Da+2A^2}>0.
\end{eqnarray}
These formulas replace Eqs. (\ref{H}), (\ref{O}), and (\ref{q}),
respectively, when the rotation differs from zero.

In the limit $a\rightarrow0$, we get again $H=1/(2t)$ and
$q\rightarrow1$, as in the standard cosmological model. In this
limit, the relative energy density takes the value $\Omega=Y_l=
1-l^2$, so that $\epsilon=3(1-l^2)/4t^2$. The expansion and
history of the primordial universe is therefore affected only by
corrections of the order of $l^2$ [see also Eq. (\ref{apsol})]. On
the other hand, in the sector of large scale factors
$a\rightarrow\infty$, one can easily check that $H=1/t$ and
$q\rightarrow0$, just like on the exact solution presented in Sec.
III. At this final stage of the expansion, the relative energy
density tends to $Y_l/(3X_l)$, a limit which is positive for
$l\in[0,1)$ and differs from the value of $1/3$, corresponding to
the non-rotating case, by terms of the order of $l^2$, supposing
that the rotation parameter is small.

Finally, it is not difficult to prove that the relative energy
density (\ref{OR}) is a strictly decreasing function of the scale
factor. Like in the model discussed in Sec. IV, $\Omega$ remains
then bounded away from zero during the whole evolution, the lower
bound being its positive limit when $a\rightarrow\infty$.
Actually, if $l\ll 1$, the energy density is always of the same
order of magnitude as the critical one. Therefore, we see that the
introduction of an anisotropic scalar field leads to a quasiflat
universe even in the presence of rotation.

\section{Validity of the approximation}
\setcounter{equation}{0}
\renewcommand{\theequation}{\arabic{section}.\arabic{equation}}

In this section we want to carry out a quantitative analysis of
the error committed in Einstein equations by identifying the
energy-momentum tensor (\ref{EMT}) with the Einstein tensor of the
RTKO metric whose scale factor is the time function (\ref{apsol}).
More specifically, we want to show that it is possible to set an
upper bound to the rotation parameter (and hence to the present
angular velocity) so that the relative error in our estimation of
the energy-momentum tensor is smaller than a fixed quantity.

For each component of the Einstein equations, we define the
relative error introduced with our approximation as the quotient
$|G_{\mu}^{\nu}-T_{\mu}^{\nu}|/\epsilon$, where $\epsilon\equiv
- T_{0}^{0}$ is the energy density of the matter content. We
want to analyze under which circumstances these relative errors
are smaller than a given number $\Delta$. Since, for any
reasonable approximation, all relatives errors should be at
least smaller than the unity, we assume that $\Delta<1$ from now
on. As we have seen, the only non-trivial components of the
Einstein equations that are exactly solved by the evolution law
(\ref{apsol}) are those corresponding to $G^{0}_{0}$ and
$G^{3}_{3}$. For the remaining components, the error is at most
of the order of $l$ when the rotation parameter is small.

Concerning our definition of relative errors, it is clear that
$\epsilon$ is the largest diagonal component of the
energy-momentum tensor. In addition, we will see below that, in
the spacetime region and range of parameters of physical interest,
the other non-vanishing component of this tensor (namely, $T^0_2$)
is also smaller than the energy density. Therefore, with our
definition, we are just comparing the errors made in the
estimation of the energy-momentum tensor with its dominant
component.

To analyze these errors, we need to deal with factors of the form
$e^{\pm x}$ that appear in most of the non-diagonal components of
the Einstein tensor, as can be seen in Eqs. (\ref{nd}). In doing
this, we will proceed as follows. Since the model is spatially
homogeneous, we can always locate the observer at the origin. From
a physical point of view, the only phenomena that can affect the
observer at a generic, present time $\eta_0$ are those that
occurred in the spacetime region that is causally connected with
him. Thus, from now on we will restrict our discussion to that
region. Let us also suppose that we are only interested in events
that happened in a certain interval of time
$\eta\in[\eta_1,\eta_0]$, with $0\leq\eta_1<\eta_0$. Although we
will make $\eta_1=0$ at the end of our calculations, we prefer to
leave this number free for the moment in order to allow for other
possibilities.

In a RTKO spacetime, one can check that the maximum absolute
value that the coordinate $x$ can take at time $\eta<\eta_0$ in
the region that is causally connected with the origin at present
is $(\eta_0-\eta)/\sqrt{1-l^2}$. A point at time $\eta$ with
this value of $x$ is connected with the origin at $\eta_0$ by
the null geodesic with vanishing $z$ and
$dy/d\eta=le^{-x}/(1-l^2)$. Hence, the region of the spacetime
that we want to analyze is contained in
\begin{equation}\label{region}
\left\{ x\in I_{\eta} \equiv
\left[\frac{-\eta_0+\eta}{\sqrt{1-l^2}},\frac{\eta_0-\eta}{\sqrt{1-l^2}}
\right],\hspace*{.4cm} \eta\in[\eta_1,\eta_0]\right\}.
\end{equation}
In particular, for each fixed value of $\eta$, the extrema of
the interval $I_{\eta}$ correspond to points that are causally
connected with the observer.

Moreover, the above region is invariant under the reversal
$x\rightarrow -x$. Using this fact and recalling that
$l\in[0,1)$ and $D\geq 0$, it is possible to show that, among
all the conditions coming from the requirement that the relative
errors be smaller than the quantity $\Delta$, the most
restrictive condition is that corresponding to the non-diagonal
component $T^2_0$ of the energy-momentum tensor. This component
leads to the inequality
\begin{equation} \label{C1}
\frac{le^{-x}}{Y_l}\frac{6X_la^2+6Da+8A^2}
{3a^2+6Da+6A^2}\leq\Delta,
\end{equation}
where $a=a(\eta)$ is given by Eq. (\ref{apsol}), the pair of
coordinates $(\eta,x)$ must belong to the region (\ref{region}),
and we have adopted again the notation (\ref{XYL}).

On the other hand, from expression (\ref{T02}), we get
\begin{equation}
\frac{|T^0_2|}{\epsilon}=le^{x}\frac{6Da+8A^2}
{3a^2+6Da+6A^2}.
\end{equation}
Recalling that the region under analysis is invariant under a flip
of sign in the coordinate $x$ (and that $X_l>0$, $Y_l\leq 1$ and
$\Delta<1$), we then see that condition (\ref{C1}) ensures that,
in the region of physical interest, the energy density dominates
over the non-diagonal component $T^0_2$ of the energy-momentum
tensor, as we had commented above.

In addition, note that, since $e^{-x}$ is a strictly decreasing
function of $x$, its maximum value for $x\in I_{\eta}$ is obtained
at $(-\eta_0+\eta)/\sqrt{1-l^2}$. So, the most stringent condition
contained in Eq. (\ref{C1}) is
\begin{equation} \label{C2}
\frac{l}{Y_l}\leq \exp{\left[\frac{\eta(a)-\eta(a_0)}
{\sqrt{Y_l}}\right]}\frac
{3a^2+6Da+6A^2}{6X_la^2+6Da+8A^2}\Delta.
\end{equation}
We have employed here relation (\ref{conformal}) to write the
conformal time in terms of $a\in[a_1,a_0]$, with $a_0>a_1$.
These two values of the scale factor are reached, respectively,
at the present time $\eta_0$ and at the initial time of our
considerations $\eta_1$.

Using the explicit form of the function $\eta(a)$, it is actually
possible to show that, for fixed parameters $l$ and $\Delta$, the
right-hand side of the above inequality is an increasing function
of $a$. As a consequence, its minimum value in the interval
$[a_1,a_0]$ is attained when $a=a_1$. In this way, we conclude
that the necessary and sufficient condition for the relative
errors to be smaller than $\Delta$ in the region of physical
relevance is obtained from Eq. (\ref{C2}) by making $a=a_1$. In
particular, if we consider the whole region that can be causally
connected with the origin since the initial big bang, i.e.
$a_1=0$, we get
\begin{equation} \label{C3}
\frac{l}{Y_l}\leq
\exp{\left[\frac{-\eta(a_0)}{\sqrt{Y_l}}\right]}
\frac{3}{4}\Delta,
\end{equation}
where we have employed that $\eta$ vanishes when $a=0$.

This inequality sets an upper bound to $l$, beyond which our
solution cannot be considered a good approximation modulo relative
errors smaller than $\Delta$. It is worth noticing that the
conformal time $\eta(a_0)$ that appears in Eq. (\ref{C3}) depends
on the rotation parameter $l$, as well as on the constants $A$ and
$D$, via relation (\ref{conformal}). Owing to this dependence, it
is in general difficult to find the exact value of the upper bound
on $l$ once the scale factor $a_0$ and the numbers $\Delta$, $A$,
and $D$ are known. In the Appendix, we present a method to
estimate such an upper bound with great accuracy. In practice,
nevertheless, it is possible to get a really good estimate by
simply replacing $Y_l=1-l^2$ with the unity and substituting
$\eta(a_0)$ by the value $\eta_0$ of the present conformal time
corresponding to the exact solution with vanishing rotation
parameter. It is not difficult to check that these approximations
amount to disregarding corrections of the order of $l^2$ in the
upper bound on $l$. Employing the values of $a_0$, $A$, and $D$
given in Sec. IV, one arrives in this way at
\begin{equation}\label{estim}
l\leq0.0337\Delta.\end{equation}

As we have said, a more careful procedure to estimate this upper
bound is presented in the Appendix, where we also consider the
possibility $a_1=a_0/1500$, corresponding approximately to the
time of decoupling between dust and radiation, and a model with
slightly different cosmological parameters, $\Omega_d=0.3$ and
$H=70$ km/(sMpc). In all these cases, we obtain a value of the
upper bound which is close to the result given above.

From inequality (\ref{estim}), we can easily derive an upper bound
on the global angular velocity at present. Using that
$\omega=l/(2a)$ and $a_0=1.2\times10^{26}$ m (the value obtained
in Sec. IV), we get $\omega\leq4.1\times10^{-20}\Delta$ s$^{-1}$.
Thus, in order to have a relative error $\Delta\leq2.5\%$ one
needs to impose, approximately, that $\omega\leq10^{-21}$
s$^{-1}$, while a more permissive error $\Delta\leq25\%$ would
lead to $\omega\leq10^{-20}$ s$^{-1}$.

Up to date, there exists no well-established and generally
accepted estimate of the angular velocity of the universe. In
models with shear, some upper bounds can be inferred from the CMB
anisotropy, but these bounds do not apply to the shear-free RTKO
spacetimes. There are some estimations of $\omega$ based on the
observed rotation of the plane of polarization of cosmic
electromagnetic radiation \cite{KO,Obukhov,Kuhne}, leading to
$\omega\sim10^{-18}$ s$^{-1}$. However, such observations are very
controversial, and the derived value of $\omega$ could well be two
or three orders of magnitude smaller \cite{Obukhov2}.

An independent estimate $\omega\sim10^{-21}$ s$^{-1}$ can be
obtained from the analysis of the rotation of galaxies \cite{Li}.
This result agrees with another estimation that is not based on
observation, but on a heuristic argument, namely, the extension to
the problem of rotation of the large number hypothesis put forward
by Dirac. The angular momentum of the observed universe is
$L\sim\rho \omega a^5$, where $\rho$ is the density of matter.
From the large number hypothesis, we get $L\sim\hbar\Lambda_D^3$
\cite{FPL}, where $\hbar$ is Planck constant and
$\Lambda_D\sim10^{39}$ is Dirac scaling parameter \cite{Dirac}.
So, we have $\omega\sim\hbar\Lambda_D^3/(\rho a^5)$. With
$\rho=3\times10^{-27}$ kg/m$^3$ and $a=a_0=1.2\times10^{26}$ m,
this leads to $\omega\sim10^{-21}$ s$^{-1}$.

Let us finally remark that the upper bound that we have obtained
for $l$ is only aimed at determining the interval of rotation
parameters in which the approximate RTKO solution presented in
Sec. V is acceptably good. In principle, rotating solutions with
larger angular velocities are possible, but their energy-momentum
tensor cannot be approximated by the matter content considered
here. On the other hand, additional restrictions on the rotation
parameter $l$ could come from the requirement that the
anisotropies that arise in the formulas of the luminosity and
angular diameter distances are compatible with the observational
data. The consideration of these anisotropies, however, cannot be
carried out analytically if $l\neq 0$, because, by contrast with
the situation found in the case with vanishing rotation (see Sec.
IV), the exact dependence of these distances with the redshift $Z$
is not manageable anymore. What is available now is (the first
terms of) their Kristian-Sachs expansion in powers of $Z$
\cite{KS}. Using the expressions given by Obukhov for this
expansion \cite{Obukhov2} and defining the relative variation of
the angular diameter distance $\varepsilon_a$ like in Sec. IV, it
is possible to show that $\varepsilon_a\simeq 2l$ up to second
order corrections in $Z$ and in the rotation parameter. Therefore,
recalling the bound on $l$ obtained above, we can affirm that the
influence of rotation in the formulas for distances is negligible,
at least as far as we do not consider sources of high redshift.
For high redshifts the Kristian-Sachs expansion is expected not to
be valid, and a more careful analysis is needed to determine the
relevance of the anisotropies.

\section{Conclusions}
\setcounter{equation}{0}
\renewcommand{\theequation}{\arabic{section}.\arabic{equation}}

In this paper, we have shown that it is possible to construct
anisotropic models that are at least compatible with the main
features of standard cosmology. In particu-lar, we have found an
exact solution of Einstein equations which describes an expanding
universe containing an anisotropic scalar field and a comoving
perfect fluid with two components: radiation and dust. The
solution is spatially homogeneous, but the sections of constant
time are anisotropic, its topology being the product of a
pseudosphere and a real line. Even so, the background radiation is
perfectly isotropic and the redshift experi-mented by any possible
emission of the dust particles varies with the scale factor like
in a FRW model. Moreover, the expansion is conformal and follows
the same evolution law as in a standard open FRW spacetime filled
with dust and radiation.

The relation between the redshift of astronomical sources and
their angular diameter (or luminosity) distance turns out to be
anisotropic, because so is the spacetime metric. However, this
anisotropy does not conflict with the current observational data,
because the corresponding variation of distances with the line of
sight in our model is not dominant compared with the systematic
and experimental errors of the measurements.

The introduction of the massless, aniso\-tropic scalar field
leaves, nevertheless, one important imprint: the energy density of
the model is of the order of the critical density at all times.
Therefore, the universe is always quasiflat. In more detail, the
relative energy density equals the unity at the initial big-bang
singularity, like in FRW cosmology, and decreases monotonically
during the whole evolution to a lower bound of one third, which is
the asymptotic limit reached at infinitely large times.

The cosmological model that we have constructed is not
completely realistic because, for instance, it does not predict
the observed accelerated expansion of the universe. In
principle, this defect could be cured by including additional
dark energy in the system, supplied either by a cosmological
constant or by a quintessence field. This modification of our
model will be discussed elsewhere. Here, we have concentrated
our attention in our simple model because it permits a clear
discussion of the effects of the anisotropic scalar field and
allows to obtain explicitly the time dependence of the scale
factor and the cosmological parameters.

We have also presented a quite straightforward application of our
exact solution, namely, the obtention of approximate cosmological
models describing spatially homogeneous, anisotropic spacetimes
with rotation. This has been possible because the anisotropic
metric of our exact solution is in fact the element with vanishing
rotation of a family of shear-free rotating metrics with
remarkable properties, including the isotropy of the comoving CMB
and the preservation of the standard relation between the redshift
of light and the value of the scale factor when this light was
emitted.

Assuming that the matter content is the same as in our exact
non-rotating solution, we have proved that it is possible to
generalize the time dependence of the scale factor so as to attain
an approximate solution of Einstein equations in the presence of
rotation. More specifically, if one restricts all considerations
to the causal past of the observer, we have shown that the error
committed with our approximations in Einstein equations, relative
to the energy density of the system (which is the dominant
component of the energy-momentum tensor), remains smaller than any
required quantity $\Delta$ if one sets an upper bound linear in
$\Delta$ to the angular velocity of the present universe. In
particular, we have calculated this bound using the values of the
Hubble parameter and the relative energy density of pressureless
matter provided by the concordance model \cite{conc}. For relative
errors of a few percent, the upper bound that we have found turns
out to be of the same order of magnitude as those obtained from
observation of the rotation of galaxies \cite{Li} and heuristic
considerations involving the large number hypothesis \cite{FPL}.

Finally, an interesting possibility would be to analyze the
angular power spectrum of primordial fluctuations in the CMB of
these anisotropic cosmologies. This analysis would require an
extension of the standard scheme of cosmological perturbation
theory \cite{cpt} that dealt with the fact that the spatial
sections of the spacetime are not maximally symmetric, took into
account the anisotropic dependence of distances on the redshift,
and treated the rotation parameter also in a perturbative manner.
These issues will be the subject of future research.

\section*{Acknowledgments}

The authors want to thank P.F. Gonz\'{a}lez-D\'{\i}az and M. Moles for
helpful comments and suggestions. S.C. was partially supported by
CNPq. G.A.M.M. was supported by funds provided by DGESIC under the
Research Project No. PB97-1218.

\section*{Appendix}
\setcounter{equation}{0}
\renewcommand{\theequation}{A\arabic{equation}}

In this appendix, we will estimate the upper bound that
inequality (\ref{C2}), evaluated at $a=a_1$, sets to the
rotation parameter $l$. Remembering expression
(\ref{conformal}), we can write the considered inequality as
\begin{equation}\label{C4}
\frac{l}{(1-l^2)}\leq J(X_l,X_l)\Delta,\end{equation} where
\begin{eqnarray}&& J(U,V)\;\;\equiv \;\;
\frac{3a_1^2+6Da_1+6A^2}{\;6Ua_1+6Da_1+8A^2\;}\nonumber \\ &&
\times\!\left(\frac{3Ua_1+D+\sqrt{9U^2a_1^2+6UDa_1+6UA^2}}
{3Ua_0+D+\sqrt{9U^2a_0^2+6UDa_0+6UA^2}}\right)^{\!\!\sqrt{2
/V}}\!\!\!\!\!.\end{eqnarray} Note that $J$ depends on the
non-negative constants $A$ and $D$ and on the values of the
scale factor at present, $a_0$, and at the initial time, $a_1$.

It is straightforward to see that, for $a_0>a_1$, $J(U,V)$
increases with $V$, assuming that $U$ and $V$ are positive. Since,
according to Eq. (\ref{XYL}), $X_l$ ranges in $(1/2,1]$, it then
follows that a necessary condition for Eq. (\ref{C4}) to be
satisfied is that $l/(1-l^2)\leq J(X_l,1)\Delta$. In addition, one
can check that $l/(1-l^2)$ is greater than $J(X_l,1)$ when $l$
approaches the unity, whereas the opposite happens at $l=0$,
provided that $A>0$. Therefore, the functions $l/(1-l^2)$ and
$J(X_l,1)$ intersect each other at least once in $l\in[0,1)$.
Moreover, in this interval of $l$, both functions turn out to be
strictly increasing. It is then possible to prove that the largest
of the intersection points,
\begin{equation}\label{lM}
L\equiv {\rm
max}\left\{l\in[0,1):\hspace*{.4cm}\frac{l}{1-l^2}=J(X_l,1)\right\},
\end{equation}
can be obtained by numerical iteration. Namely, defining $l_1=1$
and $l_{n+1}=f(l_n)$, one can get $L$ as the limit of the
sequence $\{l_n\}$, where $f(l)\equiv F[J(X_l,1)]$ and
\begin{equation}\label{Fl}
F[J]\equiv
\frac{\sqrt{1+4J^2}-1}{2J}.\end{equation}
Recalling then that $\Delta<1$ and $1-l^2\leq 1$, one easily
concludes that a necessary condition for inequality (\ref{C4})
to hold is
\begin{equation}\label{nec}
l\leq J(X_L,1)\Delta.\end{equation}

Let us now find a sufficient condition ensuring inequality
(\ref{C4}). From our previous discussion, we already know that
$l\leq L$ and that $J(U,V)$ increases with $V$ if $U$ and $V$ are
positive. Employing the definition of $X_l$, we then see that
$J(X_l,X_l)\geq J(X_l,X_L)$. In addition, $J(X_l,X_L)$ is an
increasing function of $l$ in the interval $[0,1)$, regardless of
the constant value of $X_L\in(1/2,1]$. So, $J(X_l,X_L)\geq
J(1,X_L)$, since $X_l$ becomes the unity at $l=0$. Hence, it
follows that a sufficient condition for Eq. (\ref{C4}) to hold is
$l/(1-l^2)\leq J(1,X_L)\Delta$ or, equi-valently $l\leq
F[J(1,X_L)\Delta]$. Finally, taking into account that
$0<J(1,X_L)\Delta<1$ for all the allowed values of $\Delta$ and
$X_L$, and that $F[J]\geq J(1-J^2)$ if $0\leq J\leq 1$, it is easy
to derive the simpler sufficient condition
\begin{equation}\label{suf}
l\leq J(1,X_L)[1-J^2(1,X_L)]\Delta.\end{equation}

Using the values of $a_0$, $A$ and $D$ obtained in Sec. IV, making
$a_1=0$ (i.e., considering the entire causal past of the origin),
and following the procedure explained above to determine the value
of $L$, one can check that the ne-cessary and sufficient
conditions given in Eqs. (\ref{nec}) and (\ref{suf}) lead in fact
to coincident upper bounds on $l$, up to the third significant
figure. With this degree of accuracy, one gets the bound $l\leq
0.0337\Delta$, which reproduces in fact the estimate reached in
Sec. VI. If one made instead $a_1= a_0/1500$, paying thus
attention only to those events in the causal region which occurred
(approximately) after the time of decoupling, one would obtain,
with the same level of precision, $l\leq 0.0442\Delta$.

In order to check the sensibility of our estimates to the
particular values adopted for the relative energy density of dust
matter and the Hubble parameter, we have repeated the evaluation
of the constants $A$ and $D$, the present scale factor $a_0$, and
the upper bound on $l$ taking $\Omega_d=0.3$ and $H_0=70$
km/(sMpc). In this case, following the arguments explained in Sec.
IV, one gets $A=1.3\times10^{24}$ m, $D=7.2\times10^{25}$ m, and
$a_0=1.1\times10^{26}$ m, which are close to the values found with
$H_0=65$ km/(sMpc) and $\Omega_d=0.35$. In addition, with $a_1=0$,
Eqs. (\ref{nec}) and (\ref{suf}) lead now to the bound $l\leq
0.0260\Delta$ (again up to the third significant figure), whereas
$l\leq 0.0344\Delta$ if $a_1=a_0/1500$. So, the upper bound
reached for $l$ is of the same order of magnitude in all the
considered cases.


\begin{references}

\bibitem{cobe} G.F. Smoot {\it et al.}, Astrophys. J. Lett. {\bf 396}, L1
(1992).

\bibitem{cpt} For a recent review, see, e.g., E. Bertschinger,
astro-ph/0101009 [Proceedings of COSMOLOGY 2000, edited by M.C.
Bento, O. Bertolami, and L. Teodoro (in press)].

\bibitem{Boom-Max} P. de Bernardis {\it et al.}, Nature (London)
{\bf 404}, 955 (2000); S. Hanany {\it et al.}, Astrophys. J. Lett.
{\bf 545}, L5 (2000).

\bibitem{BM2}  A. Balbi {\it et al.}, Astrophys. J. Lett.
{\bf 545}, L1 (2000);
A.E. Lange {\it et al.}, Phys. Rev. D {\bf 63}, 042001 (2001).

\bibitem{hom} R. Giovanelli {\it et al.}, Astrophys. J. Lett.
{\bf 505}, L91
(1998); D.A. Dale {\it et al.}, {\it ibid.} {\bf 510}, L11
(1999).

\bibitem{Weinberg} See, e.g., S. Weinberg, {\it Gravitation and Cosmology}
(Wiley, New York, 1972).

\bibitem{kram} D. Kramer {\it et al.},
{\it Exact Solutions of Einstein's Field Equations$\;$}
(Cambridge University Press, Cambridge, England, 1980).

\bibitem{KO} V.A. Korotki\v{\i} and Yu.N. Obukhov, Sov. Phys. JETP
{\bf 72}, 11 (1991) [Zh. Eksp. Teor. Fiz. {\bf 99}, 22 (1991)].

\bibitem{EGS} J. Ehlers, P. Geren, and R.K. Sachs, J. Math.
Phys. {\bf 9}, 1344 (1968).

\bibitem{inf} A. Guth, Phys. Rev. D {\bf 23}, 347 (1981); for a
review see D.H. Lyth and A. Riotto, Phys. Rep. {\bf 314}, 1
(1999).

\bibitem{BD} C. Brans and R.H. Dicke, Phys. Rev. {\bf 124}, 925
(1961).

\bibitem{sne} A.G. Riess {\it et al.}, Astron. J. {\bf 116}, 1009 (1998);
S. Perlmutter {\it et al.}, Astrophys. J. {\bf 517}, 565 (1999).

\bibitem{quin} J. Frieman {\it et al.}, Phys. Rev. Lett. {\bf 75}, 2077
(1995); R.R. Caldwell, R. Dave, and P.J. Steinhardt, {\it ibid.}
{\bf 80}, 1582 (1998); A.A. Starobinsky, JETP Lett. {\bf 68}, 757
(1998) [Pisma Zh. Eksp. Teor. Fiz. {\bf 68}, 721 (1998)]; T.D.
Saini {\it et al.}, Phys. Rev. Lett. {\bf 85}, 1162 (2000).

\bibitem{de} L. Amendola, Phys. Rev. Lett. {\bf 86}, 196 (2001);
A. Balbi {\it et al.}, Astrophys. J. Lett. {\bf 547}, L89 (2001).

\bibitem{bama} J.D. Barrow and J. Magueijo, Phys. Lett. B {\bf
447}, 246 (1999).

\bibitem{Obukhov} Yu.N. Obukhov, Gen. Relativ. Gravit. {\bf 24}, 121
(1992).

\bibitem{RT} M.J. Rebou\c cas and J. Tiomno, Phys. Rev. D {\bf
28}, 1251 (1983).

\bibitem{para} W. Hasse and V. Perlick, J. Math. Phys. {\bf 29},
2064 (1988).

\bibitem{PRD} S. Carneiro, Phys. Rev. D {\bf 61}, 083506 (2000);
gr-qc/0003096.

\bibitem{Landau} L.D. Landau and E.M. Lifshitz, {\it The
Classical Theory of Fields} (Pergamon, Oxford, 1983), Secs.
111-114.

\bibitem{HE} S.W. Hawking and G.F.R. Ellis, {\it The large
scale structure of space-time} (Cambridge University Press,
Cambridge, England, 1973).

\bibitem{WJJ} W. Chen and Y.S. Wu, Phys. Rev. D {\bf 41}, 695
(1990); M.V. John and K.B. Joseph, {\it ibid.} {\bf 61}, 087304
(2000).

\bibitem{conc} See, e.g., M. Tegmark, M. Zaldarriaga, and A.J.S.
Hamilton, Phys. Rev. D {\bf 63}, 043007 (2001).

\bibitem{Obukhov2} Yu.N. Obukhov, in {\it Colloquium on Cosmic
Rotation}, edited by M. Scherfner, T. Chrobok, and M. Shefaat
(Wissenschaft und Technik Verlag, Berlin, 2000).

\bibitem{KS} J. Kristian and R.K. Sachs, Astrophys. J. {\bf 143},
379 (1966).

\bibitem{MCET} K. Tomita, Prog. Theor. Phys. {\bf 40}, 264 (1968);
M.A.H. MacCallum and G.F.R. Ellis, Commun. Math. Phys. {\bf 19},
31 (1970).

\bibitem{age} J.J. Cowan {\it et al.}, Astrophys. J. {\bf 480}, 246
(1997).

\bibitem{age2} R.G. Gratton {\it et al.}, Astrophys. J. {\bf 491}, 749
(1997); B. Chaboyer {\it et al.}, {\it ibid.} {\bf 494}, 96
(1998).

\bibitem{Kuhne} R.W. K\"uhne, Mod. Phys. Lett. A {\bf 12}, 2473 (1997).


\bibitem{Li} L.X. Li, Gen. Relativ. Gravit. {\bf 30}, 497 (1998).

\bibitem{FPL} S. Carneiro, Found. Phys. Lett. {\bf 11}, 95 (1998).

\bibitem{Dirac} P.A.M. Dirac, Nature (London) {\bf 139}, 323 (1937).

\end{references}
\end{document}